\documentclass[reprint,superscriptaddress,amsmath,amssymb,aps,prl,floatfix]{revtex4-1}
\usepackage{graphicx} 

\newcommand*{\src}{MXB~1659-29}
\newcommand*{\Tc}{\tilde{T}}
\newcommand*{\Teff}{\ensuremath{T_{\!\mathrm{eff}}}}
\newcommand*{\Teffinf}{\ensuremath{\Teff^\infty}}

\newcommand*{\Qimp}{\ensuremath{Q_{\mathrm{imp}}}}

\newcommand*{\Lnu}{\ensuremath{L_\nu}}
\newcommand*{\Lin}{\ensuremath{L_{\mathrm{in}}}}
\newcommand*{\dUrca}{\ensuremath{\mathrm{dU}}}

\newcommand*{\nrml}{\ensuremath{\mathrm{nrml}}}
\newcommand*{\epsdUrca}{\ensuremath{\epsilon_{\nu,\dUrca}}}

\newcommand*{\Qunit}{\mathrm{erg\,km^{-3}\,s^{-1}}}
\newcommand*{\kB}{k_{\mathrm{B}}}
\newcommand*{\Msun}{M_\odot}

\begin{document}

\title{Rapid neutrino cooling in the neutron star MXB~1659-29}

\author{Edward F. Brown}
\email{ebrown@pa.msu.edu}
\affiliation{Department of Physics and Astronomy, Michigan State University, 567 Wilson Rd, East Lansing, MI 48864, USA}

\author{Andrew Cumming}
\email{andrew.cumming@mcgill.ca}
\affiliation{Department of Physics and McGill Space Institute, McGill University, 3600 rue University, Montreal QC, Canada H3A 2T8}

\author{Farrukh J. Fattoyev}
\email{ffattoye@indiana.edu}
\author{C.~J.~Horowitz}
\email{horowit@indiana.edu}
\affiliation{Center for Exploration of Energy and Matter and Department of Physics, Indiana University, Bloomington, IN 47405, USA}

\author{Dany Page}
\email{page@astro.unam.mx}
\affiliation{Instituto de Astronom\'ia, Universidad Nacional Aut\'onoma de M\'exico, M\'exico, D.F. 04510, M\'exico}

\author{Sanjay Reddy}
\email{sareddy@uw.edu}
\affiliation{Institute for Nuclear Theory, University of Washington, Seattle, WA 98195,USA}

\date{\today}

\begin{abstract}
We show that the neutron star in the transient system \src\ has a core neutrino luminosity that 
substantially exceeds that of the modified Urca reactions (i.e., $n+n\to n+p+e^{-}+\bar{\nu}_{e}$ and inverse) and is consistent with the direct Urca ($n\to p+e^{-}+\bar{\nu}_{e}$ and inverse) reaction occurring in a small fraction of the core. Observations of the thermal relaxation of the neutron star crust following 2.5 years of accretion allow us to measure the energy deposited into the core during accretion, which is then reradiated as neutrinos, and infer the core temperature. For a nucleonic core, this requires that the nucleons are unpaired and that the proton fraction exceed a critical value to allow the direct Urca reaction to proceed. The neutron star in \src\ is the first with a firmly detected thermal component in its X-ray spectrum that needs a fast neutrino cooling process. Measurements of the temperature variation of the neutron star core during quiescence would place an upper limit on the core specific heat and serve as a check on the fraction of the neutron star core in which nucleons are unpaired.
\end{abstract}

\pacs{97.60.Jd, 97.80.Jp, 26.60.Dd, 26.60.Kp}
\maketitle

The composition and phases of matter at low temperature and supranuclear density are fundamentally important to our understanding of QCD under extreme conditions and in astrophysics. Such matter is found in the cores of neutron stars, and observations of the surface emission of neutron stars allows us to infer their internal temperatures and hence the efficiency of neutrino cooling in their interiors. The neutrino emissivity depends on the composition of the core and whether the constituent particles are paired, forming a superfluid state \cite{yakovlev.pethick:neutron,Page2006The-cooling-of-}. Therefore observations showing that rapid neutrino cooling is operating would have important implications for our understanding of dense matter \cite{yakovlev.haensel:what,Page2006Dense-Matter-in}. 

It is useful to classify neutrino emissivities based on their temperature dependence. The first is fast, or direct Urca-like reactions ($n\to p+e^{-}+\bar{\nu}_{e}$ and inverse), for which the emissivity $\epsilon_{\rm fast} = Q_{\rm f} T^6_8$. Here we use the notation $T_{8} \equiv T/10^{8}\,\mathrm{K}$.  The coefficient $Q_{\rm f}$ is a weak function of the baryon density but depends on the phase structure: $Q_{\rm f}\simeq 10^{35}\textrm{--}10^{36}\,\Qunit$ for nucleon direct Urca in normal nuclear matter \cite{lattimer91}, $Q_{\rm f}\simeq 10^{32}\textrm{--}10^{35}\,\Qunit$ for nucleon direct Urca in the presence of a meson condensed phase \cite{Maxwell1977Beta-decay-of-p,brown88:_stran}, and $Q_{\rm f}\simeq 10^{34}\textrm{--}10^{36}\,\Qunit$ for direct Urca  reactions in quark matter \cite{Iwamoto1980Quark-beta-deca,Schafer2004Neutrino-emissi}.  The second class, intermediate in strength, are Cooper pair breaking and formation (PBF) processes, for which $\epsilon_{\rm int} = Q_{\rm i}T^7_8$ and $Q_{\rm i}\simeq 10^{28}\textrm{--}10^{29}\,\Qunit\times F_{\rm PBF}(T_c/T)$.  This intermediate process only operates in regions where $T\lesssim T_c$, $T_c$ being the critical temperature for neutron superfluidity, since $F_{\rm PBF}(T_c/T)\to 0$ otherwise \cite{flowers76:_neutr,Voskresenskii1986Neutrino-emissi,Leinson2011Neutrino-emissi}.  The third class are slow, or modified Urca-like reactions \cite{chiu64:_surfac}, for which $\epsilon_{\rm slow} = Q_{\rm s} T^8_8$ and $Q_{\rm s} \simeq 10^{28}\textrm{--}10^{29}\,\Qunit$ is again a weak function of the baryon density.

It is not known if direct Urca occurs in neutron stars because the reactions are blocked by momentum conservation unless the proton fraction exceeds a critical value, $Y_p=0.11\textrm{--}0.15$ \cite{lattimer91}. Further, pairing of protons or neutrons in the core would suppress the direct Urca reactions, so that the neutrino luminosity emitted by the direct Urca process also depends on the density dependence of the pairing energy. The classical test for the presence of enhanced cooling is to observe a sample of isolated cooling neutron stars \citep[for a catalog, see Ref.][and references therein]{Vigano2013Unifying-the-ob} and compare with theoretical cooling models \cite[for a review, see][]{yakovlev.pethick:neutron,Page2006The-cooling-of-}. Of these, some may need enhanced emission \citep{Page2009Neutrino-Emissi}. Accreting neutron stars in low mass X-ray binaries have been used to place constraints on the neutrino emissivity of the neutron star core \cite{Colpi:2001}. For transient accretors, the neutron star core comes into thermal balance between accretion-induced heating and neutrino cooling. During quiescence, when the accretion halts, the surface temperature can be measured and used to infer the core temperature. For two systems, SAX~J1808.4-3658 and 1H~1905, there are only upper limits on the thermal emission from the neutron star in quiescence \cite{Jonker2007The-Cold-Neutro,Heinke2009Further-Constra}. These upper limits are stringent enough to suggest that fast neutrino cooling is operating, but the thermal emission from these stars has not been directly detected. 

The low mass X-ray binary \src\ went into quiescence in 2001 after a 2.5 year outburst \cite{wijnands.muno.ea:burst}. Observations with \emph{Chandra} and \emph{XMM} over the next 15 years showed a declining neutron star surface temperature \cite{wijnands03:mxb1659-298,Cackett2006Cooling-of-the-,Cackett2008Cooling-of-the-,Cackett2013A-Change-in-the}. \src\ had previously been in outburst 21 years earlier \cite[see][and references therein]{Cackett2006Cooling-of-the-}. In 2015, a new outburst was reported \cite{SanchezFernandez2015}, providing a second chance to follow the cooling of the source when it goes back into quiescence. Here we show that if the outburst-quiescent cycles observed to date represent the long-term average (over the thermal timescale of the core \cite{Wijnands2013Testing-the-dee}) accretion behavior of the source, then the core neutrino luminosity must be a fast cooling process consistent with direct Urca reactions occurring in $\sim 1\%$ of the neutron star core. 

Before presenting detailed numerical calculations, we first give a basic argument. Nuclear reactions in the crust (where the density is below saturation) during accretion deposit \cite[see, e.g.,][]{Haensel2008Models-of-crust} $\approx 2\,\mathrm{MeV}$ per accreted nucleon (about $2\times 10^{18}\,\mathrm{erg\,g^{-1}}$). Over the 2.5 year outburst of \src, the neutron star accumulated a mass $\approx 8\times 10^{24}\,\mathrm{g}$; as a result, the total energy deposited by these crust reactions during the outburst is $E \approx 2\times 10^{43}\,\mathrm{erg}$. Most of this heat is conducted into the core. 
If the core is in a long-term thermal equilibrium---established over the previous $\approx 3\textrm{--}30$ outbursts for the cold core of this source---this deposited energy is radiated by neutrinos between outbursts, giving a neutrino luminosity of $2\times 10^{43}\,\mathrm{erg}/20\,\mathrm{yr}\approx 3\times 10^{34}\,\mathrm{erg\,s^{-1}}$. With knowledge of the core temperature, we can use this luminosity to constrain the neutrino emission process.

Ref.~\cite{Cumming2017Lower-limit-on-} used the observed effective temperature, after the crust had thermally relaxed, of $55\,\mathrm{eV}$ \cite{Cackett2010Continued-Cooli} to derive a redshifted core temperature $\tilde{T}= 2.5\times 10^{7}\,\mathrm{K}$. (The redshifted core temperature, denoted by $\Tc$, is the temperature measured by an observer at infinity, and is the quantity that is uniform over an isothermal core.)  For fast neutrino cooling, $L_\nu\propto \Tc^6$, we infer that 
\begin{equation}\label{e.Lnu-estimate}
L_\nu \approx  10^{38}\ {\rm erg\ s^{-1}} \Tc_8^6.
\end{equation}
The temperature in a local frame is $T = \Tc e^{-\phi}$, where $\phi \equiv \frac{1}{2}\ln{g_{00}}\approx -0.5$ is the gravitational potential found by integrating the Tolman-Oppenheimer-Volkoff equation \cite[for details, see][]{Cumming2017Lower-limit-on-}. We compare Eq.~(\ref{e.Lnu-estimate}) with the integral of the emissivity for direct Urca, $\epsdUrca \sim 10^{36}\,\Qunit T_{8}^{6}$, over a volume $V_{\dUrca}$. For direct Urca reactions, having only the innermost $\approx 3\,\mathrm{km}$ ($\lesssim 2\%$ of the volume of the core) of the neutron star above threshold is sufficient to supply this neutrino luminosity.  Conversely, an emission process that is $\lesssim 10^{-2}$ as strong as direct Urca cannot supply the required neutrino luminosity, even if it were acting over the whole core volume.

Since the inferred core temperature is small, slow reactions with rates $\propto T^{8}$, such as the modified Urca, $nn\rightarrow npe^-\bar{\nu}_e$, $np\to ppe^{-}\bar{\nu}_{e}$, and their inverses, are greatly suppressed, relative to the direct Urca rate, by a factor 
$\approx 10^{-1} (m_n/m_{\pi})^4(\kB\tilde{T}/m_{n}c^{2})^2  \simeq 10^{-7}$
at $\Tc=10^8\,\mathrm{K}$. Here $m_n$ and $m_{\pi}$ denote the neutron and pion masses. Similarly, even if relatively large regions of the star were able to sustain the intermediate class of PBF neutrino reaction reactions with $L_{\nu,\mathrm{PBF}} \propto \tilde{T}^7$ the integrated neutrino luminosities would be too small. This large difference between neutrino luminosities at the low inferred core temperature allows us to unambiguously identify the occurrence of rapid cooling in \src.  

\begin{figure}[htbp]
\includegraphics[width=\linewidth]{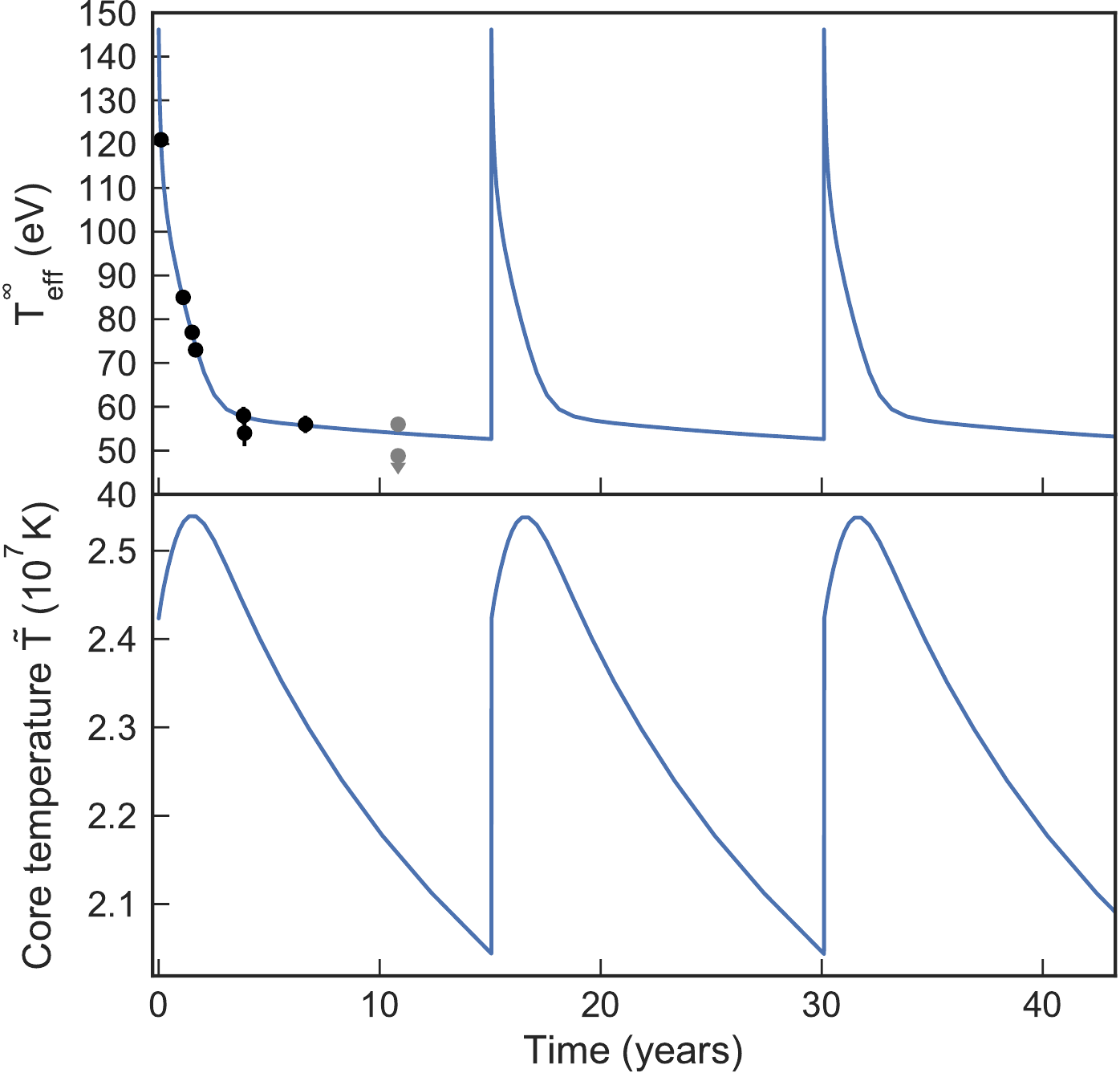}
\caption{\label{f.tc1}
An example model in long term equilibrium with core neutrino emissivity  $L_\nu=2.1\times 10^{38}\ {\rm erg\ s^{-1}}\ \tilde{T}_8^6$ and core heat capacity $C=10^{37}\ {\rm erg\ K^{-1}}\ \tilde{T}_8$. We assume identical 2.5 year outbursts  and that the source spends 5500 days ($\approx 15$ years) in quiescence. The observations are indicated as black points; in the last observation (grey point) the count rate decreased, which may be due either to increased absorption or to cooling. We assume that the neutron star temperature remains unchanged in our fits.
Other parameters are: $M=1.6\ M_\odot$, $R=11.2\ {\rm km}$, $Q_{\rm imp}=3.3$. After $\approx 2000$ days into quiescence, when the crust has come into thermal equilibrium with the core, the effective temperature continues to decline as the core cools, at a rate $\Delta \Teffinf/\Teffinf \approx 6.5\%$ per decade.}
\end{figure}

To confirm this estimate, we calculated a time-dependent model of the response of the core to many accretion outbursts. Our calculation follows the crust physics in detail \cite{Brown2009Mapping-Crustal}, but represents the core as a single zone with a total heat capacity and neutrino luminosity. This is reasonable as the thermal diffusion timescale of the core is $\ll 1\ {\rm yr}$ (rescaling eq.~[16] of Ref.~\cite{Cumming2017Lower-limit-on-} to $\tilde{T}_8=0.25$). To find the long-term steady state, we started with a guess for the core temperature just before the first outburst, ran a sequence of 50 outbursts ($\sim 1000\ {\rm yr}$) and allowed the core to come into equilibrium. Once the equilibrium core temperature was determined, we reran the sequence of outbursts with the equilibrium core temperature as a starting point. We repeated this sequence, adjusting other parameters such as the crust impurity parameter to fit the observed cooling curve for \src. Although we used a uniform outburst accretion rate and identical outburst/recurrence durations, we find that our conclusions are not sensitive to this choice \citep[cf.\ Ref.][]{Cumming2017Lower-limit-on-}.
The results are shown in Fig.~\ref{f.tc1}.
We find that a core neutrino luminosity of $2.1\times 10^{38}\ {\rm erg\ s^{-1}}\ \Tc_8^6$ provides a good fit (luminosities are given as measured by an observer at infinity). 

\begin{figure}[htbp]
\includegraphics[width=\linewidth]{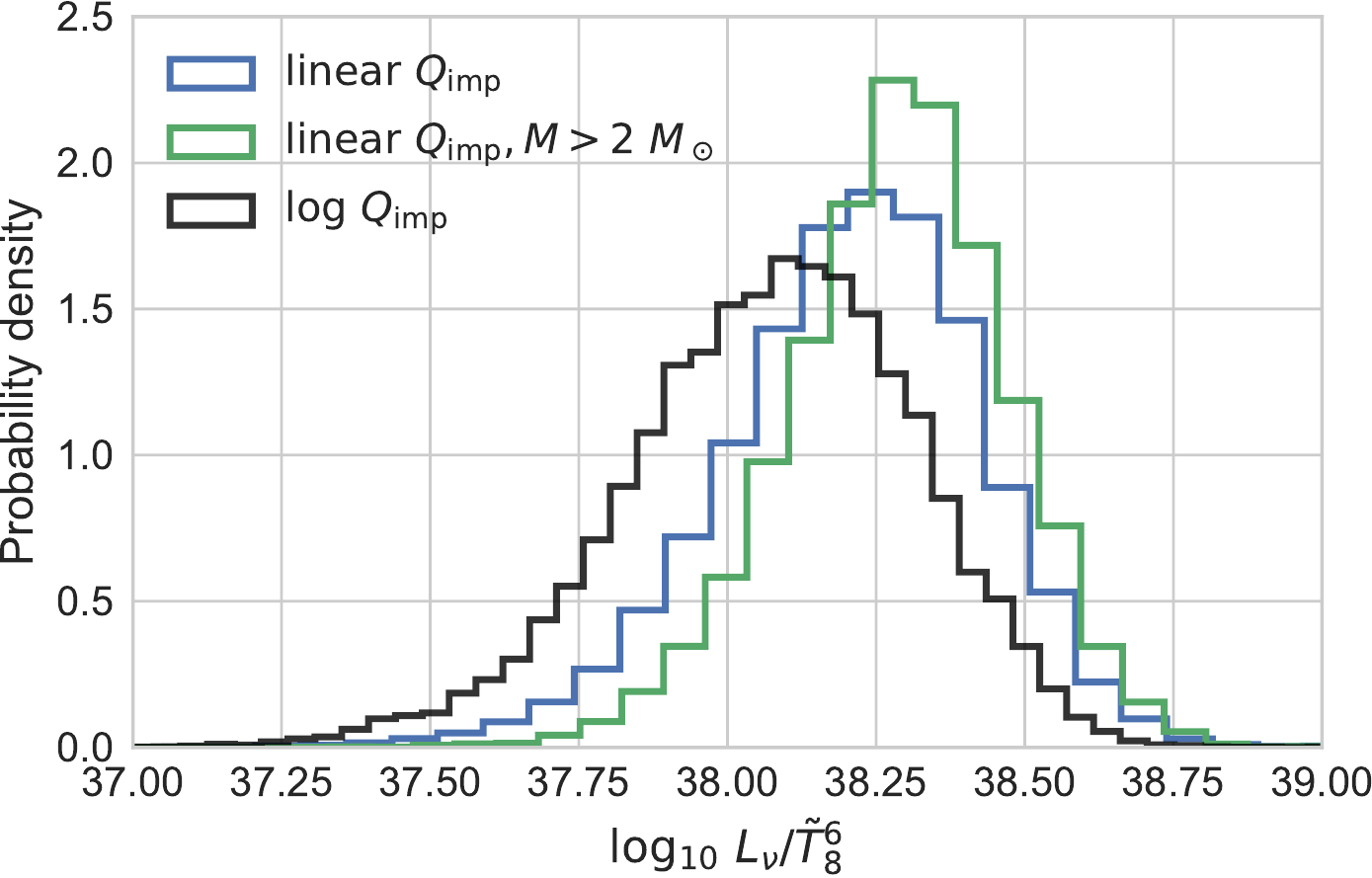}
\caption{\label{f.mcmc}
Posterior distribution of the neutrino cooling prefactor $L_\nu/\tilde{T}_8^6$ (see Eq.~[\ref{e.Lnu-estimate}]) from the MCMC fits to the \src\ cooling curve. To illustrate the robustness to different prior assumptions, we show results for three different priors: uniform in $\log_{10} Q_{\rm imp}$ (black curve), uniform in $Q_{\rm imp}$ (blue curve), and uniform prior in $Q_{\rm imp}$ with the further restriction that $M>2\ M_\odot$ (green curve).}
\end{figure}

To assess the range of allowed values for $L_\nu/\tilde{T}_8^6$, we fit the observed X-ray emission using a Markov Chain Monte Carlo algorithm. For these runs, we calculated only one outburst and kept the core temperature fixed during the simulation. By measuring the total energy conducted into the core during outburst and quiescence, we then calculated the neutrino luminosity required for the core to be in equilibrium at the assumed core temperature.  To test the robustness of our inference about $L_\nu/\tilde{T}_8^6$, we calculated three cases with different assumed priors. The resulting posterior distributions of $L_\nu/\tilde{T}_8^6$ are shown in Figure \ref{f.mcmc}. The first uses a prior for the impurity parameter \Qimp\ (a higher \Qimp\ implies a lower thermal conductivity, and hence a longer thermal relaxation, of the crust) that is uniform in $\log_{10}\,\Qimp$ (as was done in Ref.~\cite{Cumming2017Lower-limit-on-}). Because this gives much greater weight to $\Qimp\ll 1$, low values of surface gravity are preferred. While massive neutron stars (as needed to achieve the high densities required for direct Urca reactions) are allowed, they also need to have large radii to produce a low surface gravity. If instead we use a prior that is uniform in $\Qimp$, the range of surface gravity broadens and allows for more compact massive neutron stars. For the uniform prior on $\Qimp$, we find a best-fit value $\log_{10} L_\nu/\tilde{T}_8^6 = 38.2\pm 0.22$, corresponding to a central value of $1.6\times 10^{38}\,\mathrm{erg\,s^{-1}}$ with a 1-sigma standard deviation of a factor of 2 in each direction. In Fig.~\ref{f.mcmc}, the green histogram shows what happens when we impose a lower limit of $2\ M_\odot$ on the mass. In all cases, the preferred neutrino luminosity is $L_\nu/\Tc_8^6 \sim 10^{38}\,\mathrm{erg\,s^{-1}}$.

The last observation, approximately 4000 days after the end of the outburst had a decrease in the X-ray count rate. There are two possible explanations, with different implications for the inferred $\Teffinf$, as indicated by the two grey points in Fig.~\ref{f.tc1}. The neutron star surface temperature may have cooled to a lower value $<50\ {\rm eV}$ \cite{Cackett2013A-Change-in-the}.  This could be due to relaxation of the nuclear pasta layer \cite{Horowitz2015Disordered-Nucl,Deibel2017Late-time-Cooli}. The observed decrease in X-ray count rate can also be explained by an increased X-ray absorption from matter accumulating in a disk within the binary (see discussion in Ref.\ \cite{Cackett2013A-Change-in-the}). In the fits above, we assume that the neutron star temperature remained unchanged at $55\ {\rm eV}$ (upper grey point). A lower core temperature of $50\ {\rm eV}$ (lower grey point) would increase our inferred cooling rate $L/\tilde{T}_8^6$ by approximately a factor of two.

The derivation of the core temperature from the observed effective temperature depends on the composition assumed for the neutron star envelope. We have assumed a light element (He) envelope, giving a core temperature of $\tilde{T}=2.5\times 10^7\ {\rm K}$ for the observed effective temperature of $55\ {\rm eV}$. A heavy element (Fe) envelope gives a larger core temperature $\tilde{T}=5.5\times 10^7\ {\rm K}$ \cite[eq.~(5)]{Cumming2017Lower-limit-on-}, reducing the inferred value of $L_\nu/\tilde{T}_8^6$ by a factor of $\sim 100$. However, as discussed in Ref.~\cite{Cumming2017Lower-limit-on-}, the envelope composition changes the shape of the cooling curve. In the case of an Fe envelope, the hotter crust gives a much smoother decline in temperature \cite[Fig.~1]{Cumming2017Lower-limit-on-}. By varying other parameters, we can find acceptable fits to the data with an Fe envelope, but only by increasing the accretion rate (and therefore deep crustal heating rate) by a factor of $5$--$10$, which would be inconsistent with the observed persistent X-ray luminosity during outburst. In addition, a light element envelope accretes quickly ($\sim 10^{4}\,\mathrm{s}$ at typical mass accretion rates) and is therefore more likely to be left over at the end of the accretion outburst. We therefore use a light element envelope in our fits. A heavy element envelope was used in Ref.~\cite{Heinke2007Constraints-on-}, who placed \src\ in the diagram of quiescent luminosity against time-averaged accretion rate. Although our use of a light element envelope does not change the location of the contours of neutrino emissivities by a substantial amount in this diagram \cite[$\lesssim 2$; see Ref.][]{yakovlev.pethick:neutron}, it has a significant effect on the inferred value of $L_\nu/\tilde{T}^6$.

The effective temperature $55\ {\rm eV}$ was obtained assuming that the distance to \src\ is $10\ {\rm kpc}$ \cite{Cackett2010Continued-Cooli}. A 30\% larger distance \cite{Galloway2008Thermonuclear-t} gives $T_{\rm eff}$ larger  by $\approx 7$\% \cite{Cackett2008Cooling-of-the-} and $\tilde{T}$ larger by $\approx 15$\%. The resulting factor of $2$--$3$ in $\tilde{T}^6$ is offset by an increased inferred accretion rate and heating rate for a larger distance, so that overall we expect distance uncertainties to affect our inferred $L_\nu/\tilde{T}^6$ by less than a factor of 2.

\begin{figure}[htbp]
\includegraphics[width=\linewidth]{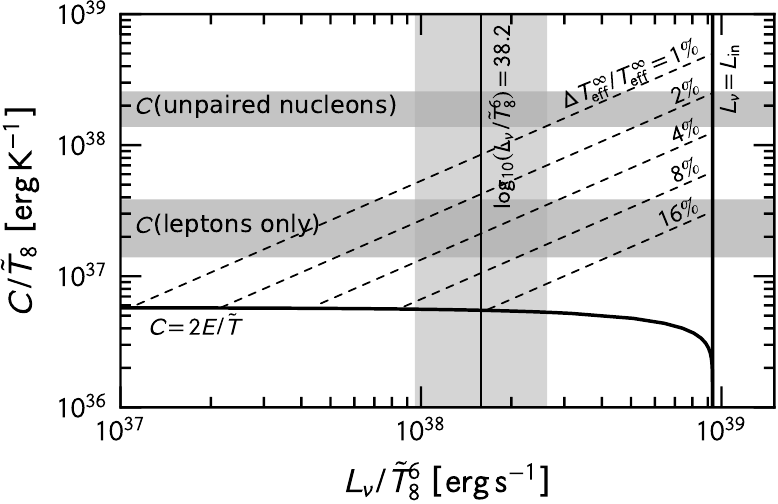}
\caption{\label{f.phaseplot}
Constraints on possible values of the specific heat $C$ and neutrino luminosity $\Lnu\propto \Tc^6$ for the $\Tc$ evaluated at the best-fit core temperature from out MCMC fits.
The lower dark curve and vertical dark line indicate the minimum specific heat and maximum neutrino luminosity compatible with $\Tc$. The vertical gray region indicates the inferred $\Lnu$. The horizontal grey regions indicate possible core specific heats. The curves marked $C = 2E/\Tc$ and $L_{\nu} = \Lin$, where $\Lin$ is the heating rate during outburst, indicate limits on the specific heat and neutrino luminosity: the neutron star lies above and to the left of these curves. The inferred $L_{\nu}$, for a uniform prior on $\Qimp$ (blue histogram, Fig.~\ref{f.mcmc}), is indicated by the vertical grey band. The plot also shows the specific heat for scenarios in which the nucleons are unpaired throughout the core (upper gray band) and in which the nucleons are fully paired, so that only leptons contribute to the specific heat (lower gray band). The bands indicate the ranges of specific heat, computed using the models in Ref.~\cite{Cumming2017Lower-limit-on-}.
The dashed lines indicate $\Delta\Teffinf/\Teffinf$ over a decade of cooling  post-outburst.
}
\end{figure}

Monitoring observations of \src\ when in quiescence, after its crust has thermally relaxed, can constrain the core specific heat. The cooling $\Delta\Teffinf/\Teffinf$ over some fixed time is $\propto\Lnu/C$ \cite{Cumming2017Lower-limit-on-} (Fig.~\ref{f.phaseplot}, dashed lines) for an inferred $\Tc$ and $E$. Figure~\ref{f.phaseplot} shows that any constraint on cooling after the next outburst will constrain the heat capacity of the core. For example, if the decrease in effective temperature, after the crust has thermally relaxed, is $\lesssim 1\%$, then a large fraction of the nucleons in the core should be normal (upper horizontal gray band), with $C/\Tc_{8}\gtrsim 10^{38}\,\mathrm{erg\,K^{-1}}$. Conversely, if the core cools significantly ($\gtrsim 4\%$) during quiescence, this would suggest that most of the nucleons are paired and the specific heat is $\lesssim 10^{37}\,\mathrm{erg\,K^{-1}}$ (lower horizontal gray band). Further modeling will be needed to separate cooling of the pasta layer \cite{Horowitz2015Disordered-Nucl,Deibel2017Late-time-Cooli} from that of the core.

The direct Urca reactions require unpaired nucleons and a sufficiently large proton fraction. The proton fraction in neutron stars is controlled entirely by the poorly known density dependence of the symmetry energy, while the critical temperature for superfluidity and superconductivity is delicate and exponentially sensitive to the details of the assumed nuclear interactions. Requiring nuclear matter to be normal constrains the heat capacity of the core, and provides a consistency check for nuclear models.  Since the volume  $V_{\dUrca}$ over which the direct Urca reaction operates is $\approx 1\%$ of the core, a large specific heat---indicated by $\Delta\Teffinf/\Teffinf\lesssim 1\%$---would imply that the volume of the core with normal matter is $V_{\nrml} \gg V_{\dUrca}$. 
Having the onset pressure be $\approx 5\%$ below the central pressure of the neutron star would generate a sufficient $V_{\dUrca}$ to supply the required neutrino luminosity.
Conversely, if $\Delta\Teffinf/\Teffinf\gtrsim 4\%$, then most of the neutrons in the star are in a superfluid phase, $V_{\nrml}\approx V_{\dUrca}$, and it is likely that the onset of direct Urca reactions is controlled by the closing of the neutron superfluid gap rather than by the proton fraction reaching the direct Urca threshold.

Assuming that fast cooling with $Q_{\mathrm{f}}=10^{36}\,\Qunit$ turns on above a given threshold density (and hence mass), \src\ is $\approx 0.03\,\Msun$ above the threshold mass, for typical equations of state. The transition may be broadened, however, over a range in density \citep[see Refs.][]{Beznogov2015Statistical-the}; alternatively, the efficiency of the direct Urca reaction could be suppressed due to many-body effects. A value of $Q_{\mathrm{f}}= 10^{35}\,\Qunit$, expected if the nucleon effective masses were significantly reduced at high density and if screening of the axial charge and two-body currents became important, would imply that the neutron star is not so close to the threshold mass. Proton pairing  may persist with a critical temperature $T_c \gtrsim 10^7\,\mathrm{K}$ over a significant volume in the core; the resulting suppression of the direct Urca rate by the factor $\simeq e^{-T_c/T}$ would favor a larger effective direct Urca volume. 

Other reactions could also be responsible for the neutrino losses, for example Urca reactions in the presence of a pion condensate or emission from unpaired quark matter.  Both of these processes can cool the star rapidly, with a rate that scales as $T^6$. The prefactor $Q_{\mathrm{f}}$ for these rates, although  poorly known, is expected to be smaller, however, than direct Urca. If $Q_{\mathrm{f}}\approx 10^{-1}$ of direct Urca then these novel phases of matter would be compatible with observations of \src. If, on the other hand, $Q_{\mathrm{f}} \approx 10^{-2}$ of direct Urca, pion condensation or unpaired quark mater are viable if most of the core were active, i.e., if the onset density for this cooling were close to saturation. Such a low onset density would imply, however, that \src\ is significantly more massive than the observed slow-cooling neutron stars. 
The neutrino luminosity of the transient neutron star KS~1731-260, for example, is $< 10^{-3}$ of the direct Urca luminosity \cite{Cumming2017Lower-limit-on-}. 
The upper limits on thermal emission from SAX~J1808.4-3658 \cite{yakovlev.pethick:neutron,Heinke2009Further-Constra,Heinke2007Constraints-on-,Beznogov2015Statistical-the,Wijnands2013Testing-the-dee} and 1H~1905+00 \cite{Jonker2007The-Cold-Neutro} imply, however, that neutron stars with much larger fast cooling volumes, and hence much higher central densities and masses, than \src\ exist. We will explore the dependence on the equation of state in a future paper. 

We have shown that with three observed, regularly spaced accretion outbursts and a measured core temperature, the neutron star in \src\  has a neutrino luminosity consistent with direct Urca reactions occurring in $\approx 1\%$ of the neutron star core. 
Taken together with KS~1731-29 \cite{Cumming2017Lower-limit-on-}, our work is further evidence that neutrino cooling in neutron star cores can be either fast or slow. Observations of the neutron star's effective temperature, now that the source has returned to quiescence, will measure the heat capacity of the core, and help to untangle the physics underlying the neutrino luminosity. The fraction of unpaired normal particles implied by the heat capacity (Fig.\ref{f.phaseplot}) will distinguish whether the direct Urca threshold is achieved only towards the center of the star, or whether superfluidity suppresses direct Urca throughout much of the core. This motivates an improved calculation of the direct Urca rate, which we shall address in a forthcoming publication.

\begin{acknowledgments}
We thank C. Heinke and W. Newton for useful discussion. This work benefited from discussions at the Physics and Astrophysics of Neutron Star Crusts workshop 2016 supported by the National Science Foundation under Grant No. PHY-1430152 (JINA Center for the Evolution of the Elements). AC is supported by an NSERC Discovery Grant and is a member of the Centre de Recherche en Astrophysique du Qu\'ebec (CRAQ). 
EFB is supported by the US National Science Foundation grant AST-1516969. FJF and CJH are supported in part by DOE grants DE-FG02-87ER40365 and DE-SC0008808. DP is partially supported by a Conacyt grant (CB-2014-1, \#240512). 
SR acknowledges support from the US Department of Energy Grant No.\ DE-FG02-00ER41132. 
We thank the Aspen Center for Physics, which is supported by National Science Foundation grant PHY-1607611, for a stimulating venue for completing this work.
\end{acknowledgments}

\end{document}